\documentclass[structabstract]{aa}

\newcommand{\saxj}{SAX~J1808.4--3658\xspace}
\newcommand{\saxjb}{SAX~J1748.9--2021\xspace}

\newcommand{\newigrj}{IGR~J17511--3057\xspace}

\newcommand{\xtejb}{XTE~J1807--294\xspace}
\newcommand{\xtejc}{XTE~J1814--338\xspace}
\newcommand{\xtejd}{XTE~J1751--305\xspace}

\newcommand{\rxte}{\textit{RXTE}\xspace}
\newcommand{\integral}{\textit{INTEGRAL}\xspace}
\newcommand{\swift}{\textit{Swift}\xspace}
\newcommand{\chandra}{\textit{Chandra}\xspace}
\newcommand{\xmm}{\textit{XMM--Newton}\xspace}

\renewcommand{\th}[0]{\ensuremath{^{\rm th}}\xspace}

\newcommand{\np}[2]{\ensuremath{#1\,\times\,10^{#2}}}

\newcommand{\nudot}{\ensuremath{\dot{\nu}}\xspace}
\newcommand{\Mdot}{\ensuremath{\dot{\rm M}}\xspace}
\newcommand{\Mdotmax}{\ensuremath{\dot{\rm M}_{\rm max}}\xspace}
\newcommand{\Mdott}{\ensuremath{\dot{\rm M}{\rm (t)}}\xspace}
\newcommand{\Tstar}{\ensuremath{T_{\star}}\xspace}
\newcommand{\Porb}{\ensuremath{P_{\rm orb}}\xspace}
\newcommand{\ecc}{\ensuremath{\it e}\xspace}
\newcommand{\Msun}{\ensuremath{{\rm M}_\odot}\xspace}

\newcommand{\changes}[1]{#1}
\newcommand{\newchanges}[1]{#1}

\usepackage{graphicx}
\usepackage{amsmath} 

\usepackage{txfonts}

\usepackage{subfigure}
\usepackage{xspace}
\usepackage{natbib}
\usepackage[english]{babel}
\usepackage{url}

\bibpunct{(}{)}{;}{a}{}{,}

\begin{document}

\title{Timing of the accreting millisecond pulsar IGR~J17511--3057}

\titlerunning{Timing of IGR~J17511--3057}

\author{A. Riggio\inst{1,2} \and A. Papitto\inst{1,2} \and
  L. Burderi\inst{2} \and T. Di Salvo\inst{3} \and M. Bachetti\inst{2}
  \and R. Iaria\inst{3} \and A. D'A\`\i\inst{3} \and
  M.T. Menna\inst{4} }

\institute{INAF/Osservatorio Astronomico di Cagliari, localit\`a
  Poggio dei Pini, strada 54, 09012 Capoterra, Italy;
  \email{ariggio@oa-cagliari.inaf.it} \and Universit\`a~di Cagliari,
  Dipartimento di Fisica, SP Monserrato-Sestu km 0,7, 09042 Monserrato
  (CA), Italy \and Dipartimento di Scienze Fisiche e Astronomiche,
  Universit\`a~di Palermo, Via Archirafi 36, Palermo, 90123 Italy \and
  Osservatorio Astronomico di Roma, Sede di Monteporzio Catone, Via
  Frascati 33, Roma, 00040 Italy}

\date{}
 
\abstract
{Timing analysis of Accretion-powered Millisecond Pulsars (AMPs) is a
  powerful tool to probe the physics of compact objects. The recently
  discovered \newigrj \changes{is the 12 discovered out of the 13} AMPs
  known. The Rossi XTE satellite provided \changes{an extensive
    coverage of the} 25 days-long observation of the source outburst.}
{Our goal is to investigate the complex interaction between the
  \changes{neutron star} magnetic field and the accretion disk,
  determining the angular momentum exchange between them. The presence
  of a millisecond coherent flux modulation allows us to investigate
  such interaction from the study of pulse arrival times. In order to
  separate the \changes{neutron star} proper spin frequency variations
  from other effects, a precise set of orbital ephemeris is
  mandatory.}
{Using timing techniques, we analysed the pulse phase delays fitting
  differential corrections to the orbital parameters. To remove the
  effects of pulse phase fluctuations we applied the timing technique
  already successfully applied to the case of an another AMP, \xtejb.}
{We report a precise set of orbital ephemeris. We demonstrate that the
  companion star is a main sequence star. We find pulse phase delays
  fluctuations on the first harmonic with a characteristic amplitude
  of about 0.05, similar to what also observed in the case of the AMP
  \xtejc. For the \changes{second} time an AMP shows a third harmonic
  detected during the entire outburst. The first harmonic phase delays
  show a puzzling behaviour, while the second harmonic phase delays
  show a clear spin-up. Also the third harmonic shows a spin-up,
  although not highly significant (3$\sigma$ c.l.). \changes{The
    presence of a fourth harmonic is also reported.} In the hypothesis
  that the second harmonic is a good tracer of the spin frequency of
  the neutron star, we find a mean spin frequency derivative for this
  source of \np{1.65(18)}{-13} Hz s$^{-1}$.}
{In order to interpret the pulse phase delays of the \changes{four}
  harmonics, we applied the disk threading model, but we obtained
  different and not compatible \Mdot estimates for each harmonic.  In
  particular, the phase delays of the first harmonic are heavily
  affected by phase noise, and consequently, from these data, it is
  not possible to derive a reliable estimate of \Mdot.  The second
  harmonic gives a \Mdot consistent with the flux assuming that the
  source is at a distance of 6.3 kpc. The third harmonic gives a lower
  \Mdot value, with respect to the first and second harmonic, and this
  would reduce the distance estimate to 3.6 kpc.}


\maketitle

\section{Introduction}
Accretion-powered millisecond pulsars (hereafter AMPs) are transient
low mass X-ray binaries, which show a coherent modulation of their
X-ray fluxes with periods of the order of few milliseconds. In the
recycling scenario AMPs are seen as the progenitors of the millisecond
radio pulsars \citep[see e.g.][]{VanDenHeuvel_84}, the accretion
process being responsible for the spinning up of the neutron star
(hereafter NS) to milliseconds periods.

The AMP \newigrj was discovered by \integral on 12 September, 2009
during a galactic bulge monitoring \citep{Baldovin_ATEL_09}.  Although
very close to the previously known AMP \xtejd, the source position
measured by \integral suggested it was a newly discovered X-ray
source. The observation of a coherent modulation of the X-ray flux in
the data from a ToO observation performed by the Rossi X-ray Timing
Explorer (hereafter \rxte) with a period of about 4 ms
\citep{Markwardt_ATEL_09} permitted to classify \newigrj as an AMP and
confirmed it as a new transient X-ray source. \cite{Altamirano_10}
reported the presence of burst oscillations at the NS frequency. An
analysis of a Chandra observation by \cite{Nowak_ATEL_09} gave the
best source position with an uncertainty of $0.6''$. \newigrj was
observed by \swift \citep{Bozzo_09}, producing a description of the
X-ray spectrum. \cite{Torres_ATEL_09} reported a possible
near-infrared counterpart. A detailed spectral analysis and a set of
orbital parameters were given by \cite{Papitto_10} analysing a ToO
\xmm observation. \cite{Riggio_ATEL_09}, analysing a \rxte
observation, refined the orbital parameters. Very recently
\cite{Miller-Jones_ATEL_09} set an upper limit on the radio
emission. Surprisingly, another transient X-ray source (\xtejd) went
into outburst very near the position of \newigrj on 7 October, 2009
\citep{Chevenez_ATEL_09} and its pulsations was detected by \rxte
\citep{Markwardt_ATEL_09b} while observing \newigrj. \newigrj has
faded under detection threshold on 8 October, 2009.

In this work we present a detailed timing analysis of the \rxte ToO
observation of the source \newigrj.

\section{Observation and Data Analysis}
In this work we analyse \rxte observation of \newigrj. In particular,
we use data from the PCA (proportional counter array) instrument on
board of the \rxte satellite (ObsId P94041 and P94042). We used data
collected in event packing mode, with time and energy resolution of
$122 \mu s$ and 64 energy channels respectively. \changes{We selected
  data in the energy range 2-25 keV in order to maximise the signal to
  noise ratio, since above $\sim$ 20 keV the background dominates.}
\begin{figure}
  \includegraphics[width=\columnwidth]{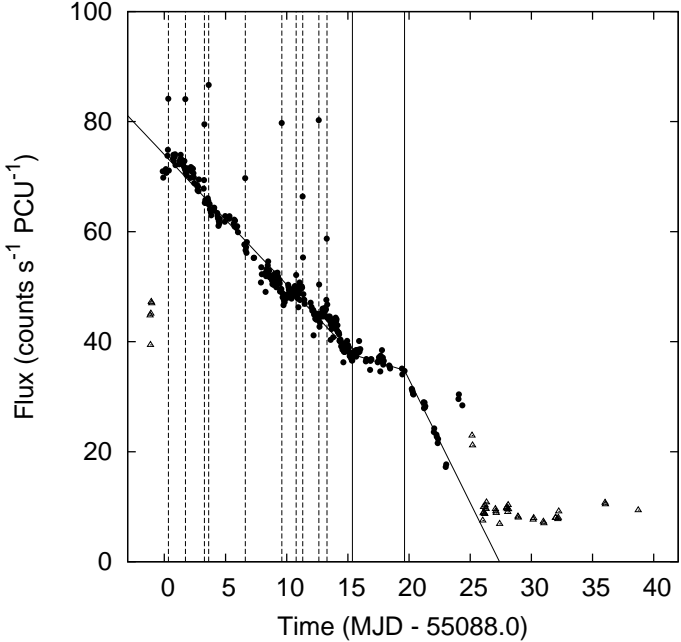}
  \caption{PCU 2 count rate \changes{(2--25 keV)}, subtracted of the
    background, is reported as a function of time during the outburst.
    The superimposed model represents the best fit using a piecewise
    linear function. The abrupt flux raise at the end of the outburst
    is due to the onset of an outburst from the AMP \xtejd. The dashed
    vertical lines are in correspondence of the type-I bursts present
    in the observation while the continuous vertical lines are in
    correspondence of the slope change accordingly to the model used
    to describe the count rate. \changes{The filled circles are
      relative to the ObsId 94041, while the triangles are the \xtejd
      observation (ObsId 94042)}}
  \label{fig:flux}
\end{figure}
The X-ray flux follows a piecewise linear decay as showed in Figure
\ref{fig:flux}, with a peak flux of 70 counts s$^{-1}$ PCU$^{-1}$.
The re-brightening visible in Figure \ref{fig:flux}, 24 days after the
start of the observation, is due to the AMP \xtejd in the field of
view of \rxte going into outburst.  The analysed data cover the time
span from 12 September 2009 (MJD 55086.8) to 22 October 2009 (MJD
55126.8). We corrected the photon arrival times for the motion of the
Earth-spacecraft system with respect to the Solar System barycentre
and reported them to barycentric dynamical times at the Solar System
barycentre using the \textit{faxbary} tool (DE-405 solar system
ephemeris).  We used the \chandra source position reported by
\cite{Nowak_ATEL_09}, and reported in Table \ref{table1}. The
uncertainty on the source position quoted by \cite{Nowak_ATEL_09} is
0.6$''$, $1 \sigma$ confidence level.

\subsection{Derivation of the orbital ephemeris}
To obtain a first estimate of the mean spin frequency we constructed a
Fourier power density spectra of the first data file of the ObsID
94041-01-01-00 with a bin size of $2^{-11}$ seconds and on data
segments of 64 seconds and averaging 53 power spectra. We found a
strong signal at $\sim 244.81$ Hz, in good agreement with the value
reported by \cite{Markwardt_ATEL_09}.
We divided the observation in time intervals of about 400 seconds each
and performed an epoch folding search on each data interval around the
averaged spin period with a period resolution of $\np{4}{-9}$ s. For
each time interval we obtained an estimate of the best spin period. We
excluded all the intervals for which the maximum in the $\chi^2$ curve
was not significant ($3 \sigma$ c.l.), according to the criterion
stated by \cite{Leahy_83}.
A sinusoidal Doppler modulation of the spin period due to the source
motion in the binary system was evident. We fitted the Doppler
frequency shifts with the formula:
\begin{equation}
  \label{eq:doppler}
  \nu(t) = \nu_0 + \dot{\nu} (t - T_0) - \frac{2\pi\, \nu_0\, A}{\Porb} \cos{l(t)},
\end{equation}
where $\nu_0$ is the spin frequency at the time $T_0$, $\dot{\nu}$ is
the spin frequency derivative, $A$ is the orbit projected semi-major
axis over the speed of light and $l(t) = 2\pi (t - \Tstar)/\Porb$,
where $\Tstar$ is the time of passage through the ascending node and
\Porb is the orbital period. With a reduced $\chi^2$(hereafter
$\chi^2_r$ and defined as $\chi^2$ / d.o.f.) of 0.59(532.2/898) we
obtained a first set of orbital parameters and a much better estimate
of the barycentric spin frequency.
\begin{table}
  \begin{minipage}[t]{\columnwidth}
    \caption{Orbital and Spin Parameters for \newigrj.}
    \label{table1}
    \centering
    \renewcommand{\footnoterule}{}  
    \begin{tabular}{lc}
      \hline \hline
      Parameter &  Value \\
      \hline
      RA (J2000), \citep{Nowak_ATEL_09}  &  \(17^{\rm h}\, 51^{\rm m}\, 08\fs66 \) \\
      Dec (J2000), \citep{Nowak_ATEL_09} &  \(-30\degr\, 57'\, 41\farcs0\) \\
      Orbital period, $P_{orb}$ (s) &  12487.5121(4) \\
      Projected semi-major axis, $a_x \sin i$ (lt-ms) &  275.1952(18) \\
      Ascending node passage, T$^\star$ (MJD) &  55088.0320279(4) \\
      Eccentricity, $\ecc$  & $<$ \np{3}{-5} \\
      Mass function$^{(1)}$, $f_x$ (\Msun) &  \np{1.070854(21)}{-3} \\
      Reference epoch, T$_0$ (MJD) & 55088.0 \\
      Mean spin frequency, $\nu_0$ (Hz) & 244.83395156(7)\\
      \hline
      \textit{Constant $\nudot$ Model best fit parameters}  & \\ 
      $\chi^2_r(\chi^2 / {\rm d.o.f.})$ & 1.74(238.5/137) \\
      Spin frequency at T$_0$,  $\nu_0$ (Hz) & 244.83395145(9) \\
      Spin frequency derivative, $\nudot$ (Hz s$^{-1}$) & \np{1.45(16)}{-13} \\
      \hline
      \textit{Physical Model best fit parameters} & \\
      $\chi^2_r(\chi^2 / {\rm d.o.f.})$ & 1.70(232.8/137) \\ 
      Spin frequency, $\nu_0$ (Hz) & 244.83395145(9) \\
      Accretion rate at T$_0$ ( \Msun year$^{-1}$)& \np{0.92(10)}{-9} \\ 
      \hline
    \end{tabular}
    \begin{flushleft}
      Errors are intended to be at $1\sigma$ c.l., upper limits are
      given at 95\% confidence level. Times are referred to the
      barycentre of the Solar System (TDB). Best fit spin parameters
      are derived in both hypothesis of a constant spin-up and flux
      dependent spin-up, and the uncertainties on the given values of
      $\nu$, \nudot and \Mdotmax include systematics due to the
      uncertainties in the source position (see text). Here we report
      only the second harmonic best fit spin parameters.\\
      $^{(1)}$ This value was obtained using the latest available
      measure of G, c (\url{http://physics.nist.gov/cuu/Constants/})
      and \Msun
      (\url{http://nssdc.gsfc.nasa.gov/planetary/factsheet/sunfact.html}).
    \end{flushleft}
  \end{minipage}
\end{table}

\begin{figure}
  \includegraphics[width=\columnwidth]{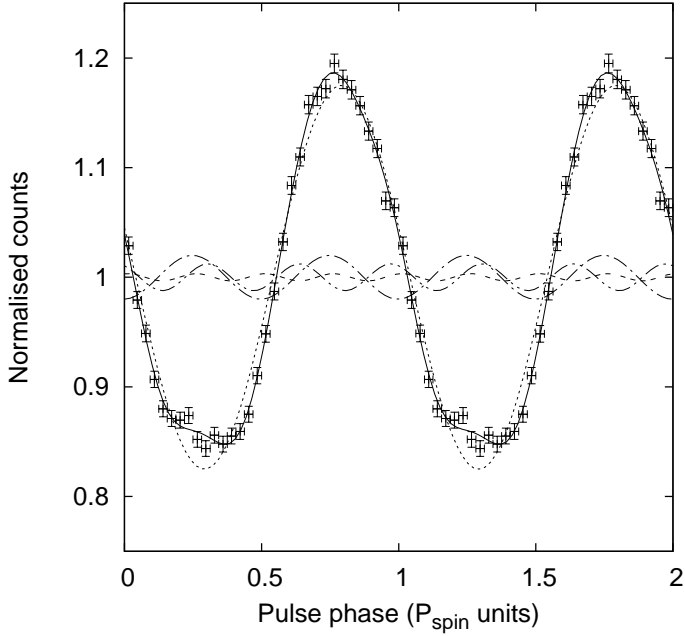}
  \caption{In this figure a folded pulse profile is reported. For
    clarity two spin cycles are plotted. The continuous line is the
    best-fit using first, second and third harmonics. We also reported
    the single contribution to the profile of the first harmonic
    (dotted line), second harmonic (dot-dashed line), third harmonic
    (bi-dot dashed line) and fourth harmonic (dashed line).}
  \label{fig:pulse_profile}
\end{figure}

\changes{Using this preliminary orbital solution} we analyse the pulse
phase delays to get a more precise estimate of the orbital and spin
parameters. We epoch folded data on time intervals of about 1500 s
using 32 phase channels. An example of the folded pulse profile is
reported in Figure \ref{fig:pulse_profile}.
An harmonic decomposition of each pulse profile up to the fourth
harmonic was necessary. To do that we fitted each normalised pulse
profile using the following expression:
\begin{equation}
  \label{eq:pulse_fit}
  f(\phi) = 1 + \sum^{4}_{n=1}{a_n \sin(2n\pi(\phi - \phi_n))},
\end{equation}
where $a_1$, $a_2$, $a_3$ and $a_4$ are the \newchanges{sinusoidal
  semi-amplitudes (hereafter fractional amplitudes)} of the first,
second, third and fourth harmonics respectively and $\phi_1$,
$\phi_2$, $\phi_3$ and $\phi_4$ the corresponding phases. We rejected
the pulse phase delays for which the following two conditions were not
both satisfied: i) the signal is not detected at least at a $3 \sigma$
confidence level \citep{Leahy_83}; ii) the best fit
\newchanges{fractional} amplitude had to be at least at 3$\sigma$ from
zero ($a_i/\delta a_i \ge 3$).

\changes{We first tried to fit the pulse phase delays with a polynomial
  to describe the pulse phase delays long term fluctuations plus the
  usual formula $\phi_{orb}(t)$, describing the pulse phase residuals
  due to differential corrections to the initial orbital parameter
  estimates\citep[][see e.g. \citealt{Riggio_07}]{Deeter_81}.}

We tried to describe the phase fluctuations using a polynomial up to
the 9\th degree to fit obtaining a $\chi^2_r = 6.11(1834.6/300)$,
which is formally unacceptable.


Due to the presence of these phase fluctuations we decided to apply
the timing technique described in \cite{Riggio_07} \changes{in order
  to} separate the orbital modulation from the phase fluctuations, to
obtain a better estimates for the orbital parameters.

Following \cite{Riggio_07} we define the pulse phase differences as:
\begin{equation}
  \label{eq:phase_diff}
  \Delta\phi_{orb}(t_i) =  \phi_{orb}(t_{i+1}) - \phi_{orb}(t_i),
\end{equation}
\changes{We excluded all the points for which $t_{i+1} - t_i > \Porb$
  in order to optimise the filter efficiency \citep{Riggio_07}.}

\begin{figure}
  \includegraphics[width=\columnwidth]{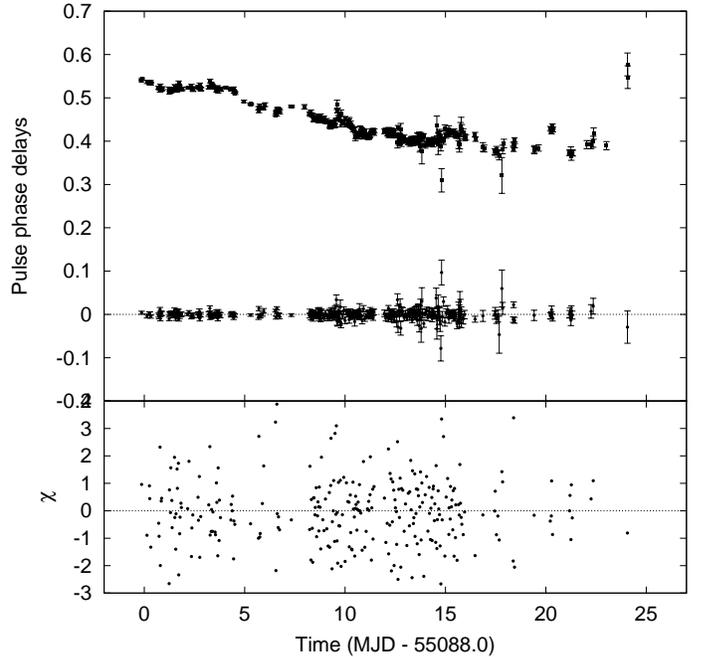}
  \caption{In the top panel of this figure the pulse phase delays
    (filled squares) and the corresponding pulse phase delay
    differences (filled circles) on the data corrected with the
    orbital ephemeris obtained from the fit of the frequency Doppler
    shifts are reported.  Each pulse phase point is obtained by
    fitting the folded pulse profile over about 1500 s of data.  It is
    evident that there is no trace of such fluctuation in the pulse
    phase delay differences, giving a striking confirmation of the
    goodness of the method. In the bottom panel are reported the
    residuals (in $\sigma$ units) with respect to the best fit orbital
    solution derived from the pulse phase delay differences using the
    timing technique described in the text.}
  \label{fig:diff_method_last_step}
\end{figure}

We iterated the process until \changes{convergence}. In the last
iteration (see Figure \ref{fig:diff_method_last_step}) the $\chi^2_r =
1.43(398/279)$, \changes{which is much more better than the previous
  approach}. The orbital ephemeris best-fit results are reported in
Table \ref{table1}, where the errors have been multiplied by the
factor $\sqrt{\chi^2_r}$ \newchanges{\citep{Bevington_03}}.

The values obtained are in perfect agreement with the ones reported by
\cite{Papitto_ATEL_09} from an analysis of a \xmm observation and by
\cite{Riggio_ATEL_09}, obtained with \rxte data but on a shorter time
interval. The obtained uncertainties on the orbital parameters are a
factor two smaller than the best orbital solution previously given by
\cite{Riggio_ATEL_09}. Note also that the orbital period we derive
here has a relative uncertainty as small as 0.03\%.

In order to fit the pulse phase delays with a physically meaningful
torque which takes into account the decreasing X-ray flux (and hence
\Mdot, in the hypothesis that the X-ray flux is a good tracer of the
mass accretion rate onto the NS), we need to describe the flux
evolution during the outburst, i.e.\ the bolometric flux at the peak
of the outburst and the shape of the light curve.

In order to estimate the bolometric flux of the source at the peak of
the outburst, we considered the spectrum collected by the PCA in
Standard 2 Mode (129 energy channels recorded at a time resolution of
16s) during Obs. 94041-01-01-04, which started on 55089.283 MJD and
had a total exposure of 10480s. We considered only data taken by the
PCU2 \changes{in order to avoid cross-calibration problem between
  PCUs. We selected only events detected} in its top Xenon layer to
maximise the signal to noise ratio\citep{Jahoda_06}. The background
was modelled using the bright source model, that is appropriate for
sources emitting $>40$ c s$^{-1}$ PCU$^{-1}$. We used the latest
version (11.7) of the PCA response matrix generator, restricting to
the 3--50 keV band and adding a systematic error of 0.5\% to spectral
counts\footnote{N. Shaposhnikov, K. Jahoda, C. B. Markwardt, 2009,
  \url{http://www.universe.nasa.gov/xrays/programs/rxte/pca/doc/rmf/pcarmf-11.7/}}.
Similarly to what was observed from other AMPs, the X-ray emission of
\newigrj is dominated by a power law--like emission extending to high
($>50$ keV) energies. To model it we considered a simple thermal
Comptonisation model (\texttt{nthcomp}, \citealt{Zdziarski_96};
\citealt{Zycki_99}), fixing the temperature of the hot electrons to
100 keV.  Given the poor coverage of the PCA at low energies, we have
also constrained the absorption column to nH = \np{1}{22} cm$^{-2}$,
as it is suggested by an \xmm observation performed during the same
outburst \citep{Papitto_10}. A 6.6 keV emission line was also added to
model residuals in the iron range, though such a feature is probably
due to the contamination of the Galactic ridge in the field of view of
the PCA \citep{Markwardt_ATEL_09}. The reduced chi square of the fit
is good ($\chi^2_r= 64.3/72$). The unabsorbed flux we detect in the
3-50 keV band is \np{1.14(1)}{-9} erg cm$^{-2}$ s$^{-1}$, that
extrapolated in the 0.5-200 keV band gives an estimate of $\sim
\np{2}{-9}$ erg cm$^{-2}$ s$^{-1}$.  Under the hypothesis that
emission is isotropic this corresponds to a bolometric luminosity of
$L_x\sim1.5\times10^{37}$ d$_8^2$ erg s$^{-1}$, where d$_8$ is the
distance to the source in units of 8 kpc. We note here that an upper
limit of $10.6$ kpc on the source distance was set by \cite{Bozzo_09},
by imposing that the burst peak luminosity does not exceed the
Eddington limit, while \citet{Altamirano_10}, from the analysis of the
type-I bursts observed by \rxte and \swift, found an upper limit of
6.9 kpc. \changes{\cite{Papitto_10}, on the basis of the spectral
  analysis of \xmm data, gave also a lower limit of $\simeq 7$ kpc,
  although derived under some assumptions.}  Moreover, as the source
is only a few degrees away from the galactic centre and its X-ray
emission is not heavily absorbed (nH $\sim 10^{22}$ cm$^{-2}$), it is
highly probable that the distance does not exceed 8 kpc.  Assuming
that $L_X=\epsilon GM\Mdot/R$ with $\epsilon \simeq 1$, we eventually
deduce a peak mass accretion rate of the order of $\simeq
\np{1.5}{-9}$ d$_8^2$ \Msun yr$^{-1}$, which is the estimate we use in
the following to compare the dynamical estimates of \Mdot from the
timing analysis.

To describe the light curve shape, we chose to fit it with a piecewise
linear function composed of three segments, as shown in Figure
\ref{fig:flux}. We choose to fit only the first 23 days of data since
the subsequent data are affected by the concomitant \xtejd outburst.
We modelled each of the three intervals with a function $c_i(t) = c_i
(1 - (t - T_i) / \tau_i), T_i \le t < T_{i+1}$, where $c_i$ is the
count rate at $t = T_i$ and $\tau_i$ the linear decay timescale for
the i-th piece. For the sake of simplicity we wrote this piecewise
function as $c_0 f(t)$, where $c_0$ is the count rate at the peak. The
best-fit result is reported in Figure \ref{fig:flux}.

\subsection{Timing analysis}
The spin frequency evolution in AMPs is thought to be driven by the
accretion process. Matter falling from the accretion disk onto the NS
transfers its angular momentum to the NS, which is spun-up to
millisecond spin periods.  But, as was evident from the first attempts
\citep{Ghosh_77}, the magnetic field - accretion disk interaction can
exert a negative torque onto the NS, spinning it down. This is called
the threaded disk model. Due to the complexity of the problem, the
details of the NS magnetosphere - disk interaction are still not well
understood. In Literature \changes{three} examples of AMPs which
spin-down while accreting are reported \citep{Galloway_02, Burderi_06,
  Papitto_08}.

As already observed in other two AMPs (see e.g. \citealt{Burderi_07}
for \saxj and \citealt{Riggio_08} for \xtejb), also for this source
the first harmonic is dominated by fluctuations and then unusable for
our scope. It should be noted that the two AMPs cited above (\saxj and
\xtejb) show a second harmonic with a more regular behaviour.

\changes{An alternative interpretation of the pulse frequency
  derivatives was given by \citet{Hartman_08}, who suggested that the
  red timing noise affecting the pulse phase delays can mimics a spin
  frequency derivative. \citet{Patruno_09} try to demonstrate that the
  pulse phase delays are correlated with the X-ray flux, rather than
  the genuine spin evolution of the source, due to motion of the
  hot-spot related to the flux. Unfortunately this correlation is not
  clear, even in the sign, in all AMPs and differs, in the same
  sources (see e.g. \saxj and \xtejb), for each harmonic component, as
  noted by \citet{Patruno_09}. We tested this hypothesis using the
  method as described in \citet{Patruno_09}, adopting a constant spin
  frequency model to derive the pulse phase residuals. In the best-fit
  which maximise the linear correlation between phase residuals and
  flux we obtained $\chi^2_r \simeq 23$ \newchanges{(4625 / 201
    d.o.f.)}, indicating that, for this source, the pulse phase
  residuals of the first harmonic cannot be ascribed to flux variation
  and/or fluctuations. }


In the following, \changes{we will work under widely accepted
  hypothesis that the pulse frequency is the NS spin frequency}. We
will analyse the pulse phase delays and apply to them a disk threading
model to derive the \Mdot and compare it with the value obtained from
the spectral analysis of the same data.

We epoch folded data on time intervals of about 3.0 ks (one pulse
profile per data file) and 32 phase bins. A third harmonic is detected
during all the outburst. A fourth harmonic was also detected in the
first 15 days of the outburst.  In Figure \ref{fig:first_fit},
\ref{fig:snd_fit}, \ref{fig:trd_fit} and \ref{fig:forth_fit} the
first, second, third and fourth harmonic pulse phase delays are
reported, respectively. \changes{The fractional amplitudes,
  \newchanges{as defined in Eq. \ref{eq:pulse_fit}}, were corrected to
  take into account the instrumental background ($\sim 11$ counts
  s$^{-1}$ PCU$^{-1}$ in the 2-25 keV energy band) and the background
  due to the presence of the galactic ridge in the field of view of
  \rxte \citep{Markwardt_ATEL_09}. To estimate this supplementary
  background we used the observations of the AMP \xtejd when both
  sources went to quiescence, in particular the observations from MJD
  55115.400 to 55126.745, with a total exposure of 33.7 ks. Due to the
  low count rate, we used the faint background model, obtaining a
  count-rate of $\sim 7$ counts s$^{-1}$ PCU$^{-1}$ in the same energy
  band.}  \changes{In Figure \ref{fig:amplitudes} the fractional
  amplitudes for the four harmonics are reported.}

The mean spin frequency reported in Table \ref{table1} is obtained by
fitting the first harmonics pulse phase delays with a constant spin
frequency model. The value obtained in this way for the spin frequency
is $\nu = 244.8339515569(24)$ Hz. However, the systematic effects due
to the uncertainty of 0.6$''$ \citep{Nowak_ATEL_09} on the source
position brings the mean spin frequency error to \np{7}{-8} Hz
\citep[see][]{Burderi_07}.

\subsubsection{First harmonic}
We started fitting the first harmonic pulse phase delays with a
constant spin frequency derivative in order to have an estimate of the
mean spin frequency derivative. From the fit we obtained $\nudot =
\np{-7.0(9)}{-14}$ Hz s$^{-1}$ with a $\chi^2_r = 18.51(3869/209)$,
clearly unacceptable.

In order to improve this result we used the threaded disk model
applied to the AMPs by \cite{Rappaport_04}. According to this model,
the net torque acting on an AMP is:
\begin{equation}
  \tau(t) = 2 \pi I \nudot(t) = \Mdott\sqrt{GMR_c} -
  \frac{\mu^2}{9R_c^3},
  \label{eq:eqn2}
\end{equation}
where I is NS moment of inertia, \Mdot is the mass accretion rate, $M$
is the mass of the NS, $\mu$ is the magnetic dipole moment of the NS
and $R_c$ the co-rotation radius.

To apply this model to our data we need an expression for \Mdott. We
then made the hypothesis that the bolometric luminosity L(t) is a good
tracer of \Mdot. In the hypothesis that the spectral variation during
the outburst is not significant it is possible to consider L(t)
proportional to the background subtracted count rate. We can then
write $\Mdott = \Mdotmax f(t)$, where \Mdotmax is the maximum
accretion rate in correspondence of the flux peak and f(t) is the
functional form of the count rate previously derived.

An expression for \nudot as function of \Mdot can easily be derived
from Eq. \ref{eq:eqn2}:
\begin{equation}
  \label{eq_nudot}
  \nudot(t) = \Biggl[1.427\;\frac{m^{2/3} P^{1/3}_{-3}
    \dot{\rm M}_{^{\rm max}_{-10}}}{I_{45}} f(t)- \\
  5.232\; \frac{\mu^2_{26}}{mI_{45}P^2_{-3}} \Biggr] \times 10^{-14}
  \;\textrm{Hz s}^{-1},
\end{equation}
where m is the NS mass in units of \Msun, $\dot{\rm M}_{^{max}_{-10}}$
is the maximum mass accretion rate in units of $10^{-10}$ \Msun
y$^{-1}$, P$_{-3}$ the spin period in units of $10^{-3}$ s and
$I_{45}$ the NS moment of inertia in $10^{45}$ gr cm$^2$. In this work
we adopted the FPS \citep[Friedman Pandharipande Skyrme, see
][]{Friedman_81, Pandharipande_89} equation of state for which, fixing
NS mass to $M = 1.4 \Msun$, we obtain a radius of $R_{NS} = 1.14
\times 10^6$ cm and a moment of inertia $I = 1.29 \times 10^{45}$ gr
cm$^2$.

The pulse phase delays formula used for the fit is obtained doubly
integrating the expression \ref{eq_nudot} with respect to the time
\citep[see e.g.][]{Burderi_07}.

\begin{figure}
  \includegraphics[width=\columnwidth]{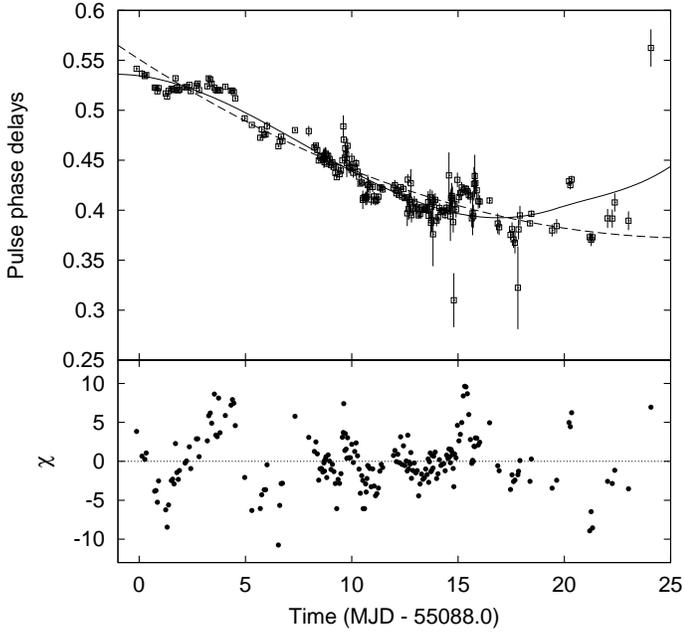}
  \caption{In the top panel we show the first harmonic pulse phase
    delays and the best-fitting curves, considering a constant spin
    frequency derivative (dashed line) and the disk threading model
    proposed by \cite{Rappaport_04}. In the bottom panel we show the
    residuals (in $\sigma$ units) of the first harmonic with respect
    to the best fitting model given by the threaded disk.}
  \label{fig:first_fit}
\end{figure}

From the fit of the first harmonic we obtained a magnetic dipole
strength of $\mu = $ \np{1.64(7)}{27} G cm$^3$ and \Mdotmax =
\np{5.7(6)}{-9} \Msun y$^{-1}$, with a $\chi^2_r = 12.36(2570/208)$,
still unacceptable. The results are reported in Figure
\ref{fig:first_fit}.
Such a big $\chi^2_r$ is clearly due to fluctuations in the pulse
phase delays, as it is possible to see in the best-fit residuals
reported in Figure \ref{fig:first_fit} (bottom panel). \changes{ As
  foretold, we unsuccessfully try to interpret it with the model
  suggested by \citet{Patruno_09}. The origin of such fluctuations
  still remains unexplained.}

\subsubsection{Second harmonic}
\begin{figure}
  \includegraphics[width=\columnwidth]{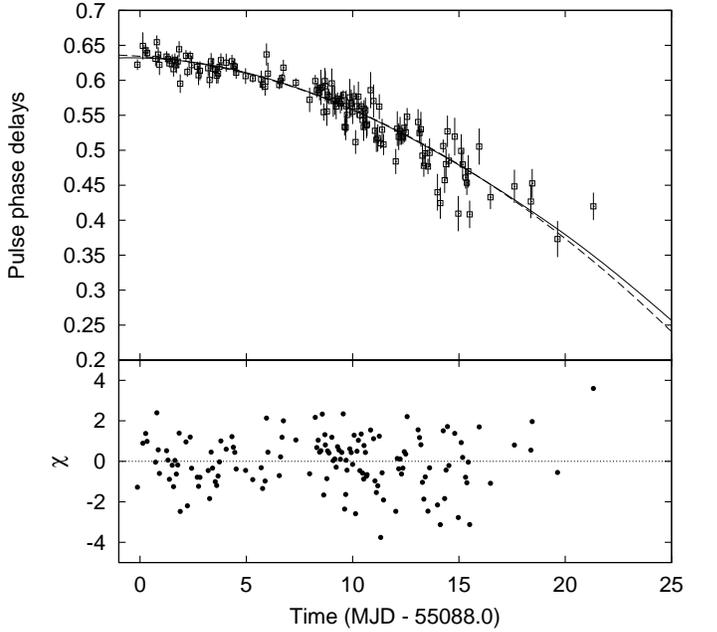}
  \caption{In the top panel of this figure the second harmonic pulse
    phase delays are shown. The dashed and the continuous lines are
    the second harmonic best-fit curves considering a constant spin
    frequency derivative and the physical model considering the
    material torque proportional to the flux, respectively. In the
    bottom panel the residuals (in $\sigma$ units) of the second
    harmonic with respect the constant spin frequency derivative are
    shown.}
  \label{fig:snd_fit}
\end{figure}

\changes{Considering the second harmonic less affected by phase noise
  \citep{Burderi_06, Riggio_08}, we repeated the fitting procedure
  using these phase delays. In the constant spin frequency derivative
  case} we obtained $\nudot = \np{1.45(16)}{-13}$ Hz s$^{-1}$ with a
$\chi^2_r = 1.74(238.5/137)$. The best-fit curve is reported in Figure
\ref{fig:snd_fit}.  We adopted the same disk threading expression used
for the first harmonic to describe the second harmonic. In this case
we had to fix $\mu = 0$ since $\mu$ and \Mdotmax strongly
correlate. In this case, therefore, the derived value of \Mdotmax has
to be considered as a lower limit to the mass accretion rate at the
peak of the outburst, since value of the magnetic moment higher than
zero will give a higher value for \Mdotmax.
The best-fit results are reported in Table \ref{table1}. In this case
the best-fit value of \Mdotmax = \np{0.92(10)}{-9} \Msun year$^{-1}$,
with a $\chi^2_r = 1.70(232.8/137)$, is in good agreement with our
estimate from the bolometric flux for a source distance of about
$6.3(3)$~kpc.

\subsubsection{Third harmonic}
\begin{figure}
  \includegraphics[width=\columnwidth]{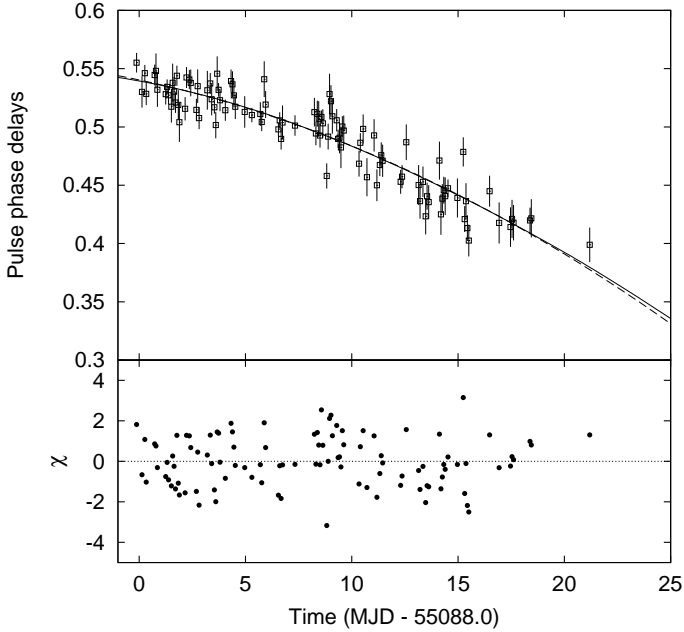}
  \caption{In the top panel of this figure the third harmonic pulse
    phase delays and its best-fit curves considering a constant spin
    frequency derivative (dashed line) and the physical model
    considering the material torque proportional to the flux
    (continuous line) are reported. In the bottom panel we show the
    residuals in units of $\sigma$ with respect the constant spin
    frequency derivative model.}
  \label{fig:trd_fit}
\end{figure}

Since in this source the third harmonic is significantly detected in
nearly the whole outburst we proceeded with the same method used for
the first and second harmonic. From the fit with the constant spin
frequency derivative we obtained $\nu = 244.83395151(6)$ Hz and
$\nudot = \np{4.8(1.4)}{-14}$ Hz s$^{-1}$ with a $\chi^2_r =
1.58(159.8/101)$. The best-fit curve is reported in Figure
\ref{fig:trd_fit}.  We applied the threaded disk model to describe the
third harmonic and, as already done for the second harmonic, we fixed
$\mu = 0$ since $\mu$ and \Mdotmax correlate in the fit.
The obtained best-fit values are \Mdotmax = \np{3.1(9)}{-10} \Msun
year$^{-1}$ for the peak mass accretion rate and $\nu =
244.83395150(6)$ Hz with a reduced $\chi^2$ of $\chi^2_r =
1.57(159.0/101)$. The third harmonic shows a spin-up, although not
highly significant ($\gtrsim 3 \sigma$ c.l.). The peak mass accretion
rate deduced is quite low in comparison to the one obtained from the
bolometric flux for a source distance of $6.3(3)$ kpc. \changes{In fact
  the source distance is of 3.6(5) kpc}.

\subsubsection{Fourth harmonic}
\begin{figure}
  \includegraphics[width=\columnwidth]{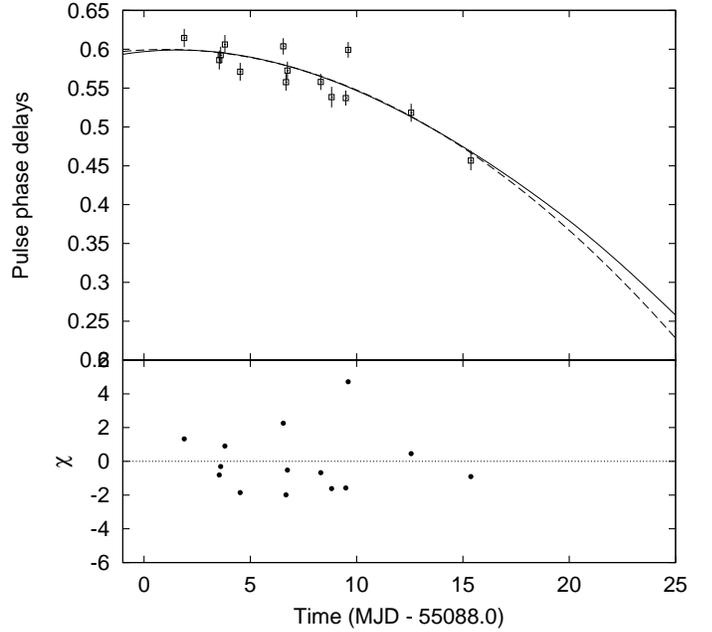}
  \caption{In the top panel of this figure the fourth harmonic pulse
    phase delays and its best-fit curves considering a constant spin
    frequency derivative (dashed line) and the physical model
    considering the material torque proportional to the flux
    (continuous line) are reported. In the bottom panel we show the
    residuals in units of $\sigma$ with respect the constant spin
    frequency derivative model.}
  \label{fig:forth_fit}
\end{figure}

\changes{The fourth harmonic is sporadically detected in the first 15
  days of the outburst, with a fractional amplitude of $\sim$
  1\%. From the fit with the constant spin frequency derivative we
  obtained $|\nudot| < \np{2.7)}{-13}$ Hz s$^{-1}$ (2$\sigma$ c.l.)
  with a $\chi^2_r = 4.03(44.28/11)$. The best-fit curve is reported
  in Figure \ref{fig:forth_fit}.}
\changes{The obtained best-fit values are $\Mdotmax < \np{1.6}{-9}$
  \Msun year$^{-1}$ (2$\sigma$ c.l.) for the peak mass accretion rate
  with a reduced $\chi^2$ of $\chi^2_r = 4.09(45/11)$.}

\vskip 1.5em 

It should be noted that, as already done in \cite{Burderi_07} and
\cite{Riggio_08}, we taken into account the effect of the source
position uncertainty on the obtained values of $\nu$ and \nudot,
adopting the same method described in \cite{Riggio_08}.  
\begin{figure}
  \includegraphics[width=9.0cm]{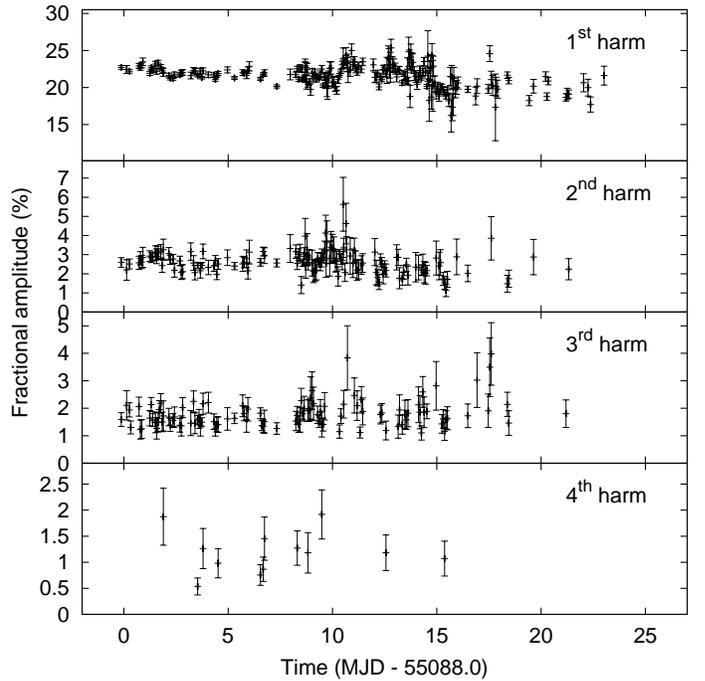}
  \caption{The fractional amplitudes of the four harmonics are
    reported. Points in correspondence of the type-I bursts are not
    plotted due to difficult estimate of the persistent flux.}
  \label{fig:amplitudes}
\end{figure}
In particular, in the case of constant spin frequency derivative the
uncertainties are of $\Delta\nu = \np{6.1}{-8}$ Hz on the frequency
and $\Delta\nudot = \np{0.72}{-14}$ Hz s$^{-1}$ on the spin frequency
derivative, while in the case of the physical model the uncertainties
are of $\Delta\nu = \np{6.1}{-8}$ Hz on the frequency and $\Delta\Mdot
= \np{4.3}{-11}$ \Msun y$^{-1}$ on the peak accretion rate.

\begin{figure*}[ht]
  \subfigure[]{\includegraphics[width=9.2cm]{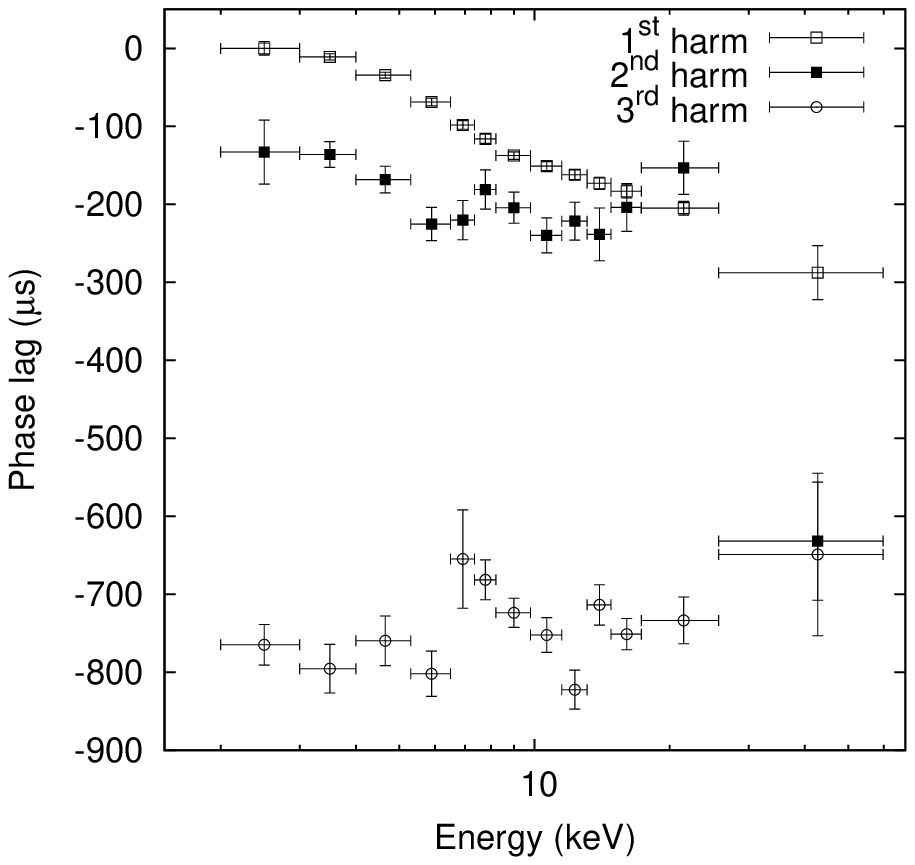}}
  \subfigure[]{\includegraphics[width=9.0cm]{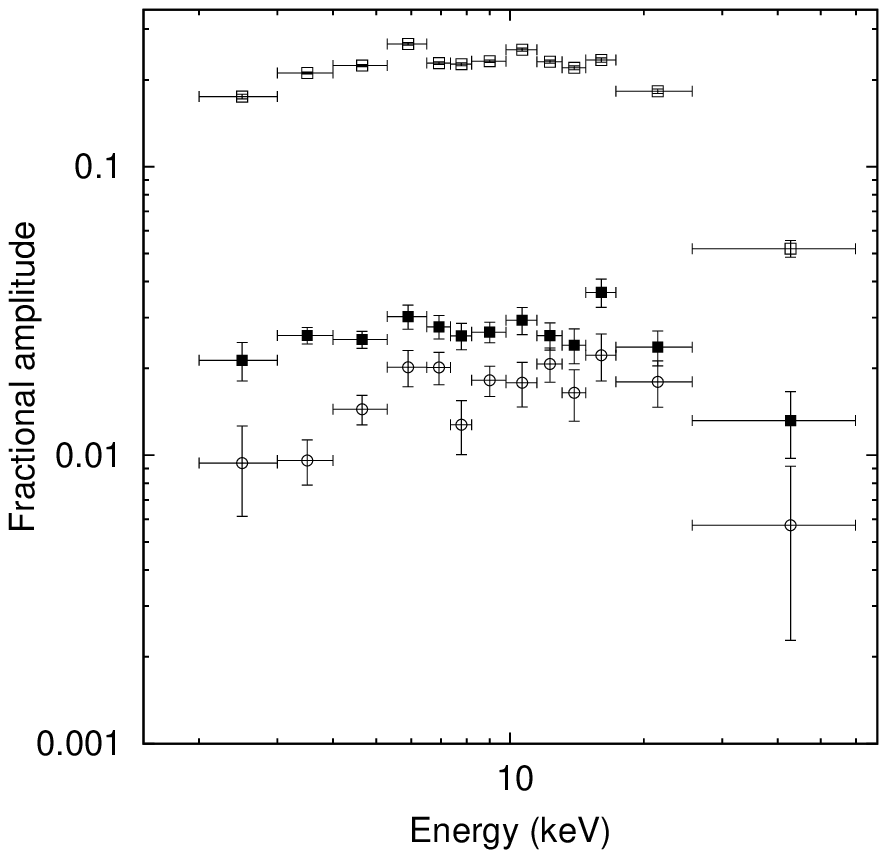}}
  \caption{In the left panel the phase lags for the first three
    harmonics are reported. The phase lags are measured with respect
    to the maxima of each component, as done in \citet{Papitto_10}. In
    the right panel the fractional amplitude for the three harmonics are
    reported.}
  \label{fig:phase_lag}
\end{figure*}

\subsection{Spectral variability of the pulse profile}
\changes{As already done by \citet{Papitto_10} analysing the \xmm
  observation of the same outburst, we analysed the energy dependence
  of phase and amplitude of the three harmonic components. In order to
  have a good statistics we epoch folded data from 55089.233 MJD to
  55092.5379 MJD, excluding the three data files where a type-I burst
  was present, with an exposure of 81 ks and a coverage of 28\%. We
  choose this interval because the pulse phase delays have a quite
  stable linear trend with an average first harmonic phase scattering
  of $0.005$. We consider only data from PCU2.}  \changes{ We
  evaluated also the contribution to the background given by the
  galactic bulge, using the data described in the previous section, in
  the working hypothesis that the changes due to the different
  observation data and slightly different instrument pointing are
  small \citep[see][for a detailed discussion]{Papitto_10}. For these
  reasons small systematics could be present and the confidence
  intervals could be underestimated.}  \changes{ The fractional
  amplitude of the first harmonic (Fig. \ref{fig:phase_lag}, right
  panel) increases from 17.5(3)\% at 2.5 keV to 26.7(3)\% at 6 keV. It
  remains roughly constant around 23\% up to 12 keV, where the
  fractional amplitude is 25.4(3)\%. Above 12 keV there is a steady
  decline of the fractional amplitude, although up to the energy band
  25-60 keV is still well detectable with a fractional amplitude of
  5.2(3)\%. The second and third harmonic fractional amplitudes show a
  behaviour similar to the first harmonic one, increasing with energy
  with a maximum (around 10-20 keV) of 3.7(4)\% and 2.2(4)\%,
  respectively. We also observe a decline of the fractional amplitude
  above 20 keV for these harmonics.  Fractional amplitudes as a
  function of photon energies were not clearly detected for the fourth
  harmonic, probably due to the long time of integration and the
  smearing caused by the observed fourth harmonic phase fluctuations
  of $\sim 0.05$ (see Fig. \ref{fig:forth_fit}), which is a
  considerable fraction of the fourth harmonic period.}


\changes{The first harmonic show phase lags (reported in
  Fig. \ref{fig:phase_lag}, left panel). As described in
  \citet{Papitto_10}, there is a steady decrease of the phase lag up
  to $\sim$ 10 keV, where it is clearly visible a break. Beyond 10 keV
  the pulse phase lag still decrease, but with a smaller rate.}

\changes{The second harmonic shows a different trend with respect to
  the first harmonic. It reaches the maximum lag around 10 keV, it
  shows no time lag at 25 keV, but then we observe a sudden jump of
  $\sim -450 \mu$s from 25 to 50 keV.}

\changes{The third harmonic phase lags are roughly constant in all the
  energy band. It should be noted that in the energy band 25.7--59.8
  keV the phase lags for the second and the third coincide.}

\subsection{type-I bursts timing}
We performed a timing analysis of all the 10 type-I burst present in
the observation, with the goal of studying the pulse profile evolution
during the burst. In Figure~\ref{fig:burst_ev} we show the results for
the second and fifth type-I burst, the best sampled among the bursts
present in the \rxte observation. The starting date of these bursts is
55089.721 MJD (TDB) and 55094.619 MJD (TDB), and the decay time is
8.0(1) seconds and 8.5(1) seconds, respectively.  We divided each
burst in chunks holding (roughly) the same number of events so that in
each folded profile a fractional amplitude of $\sim 20\%$ can be
easily detectable. We folded each chunk using 8 phase bins and
performed an harmonic decomposition using only the first harmonic.
The results of this analysis are reported in Figure
\ref{fig:burst_ev}. During both bursts, the pulse phase delays remain
stable and, with the exception of the very first seconds, locked to
the pulse phase delays during the persistent emission (see Figure
\ref{fig:burst_ev}, mid panel). The fractional amplitude behaviour is
even more interesting because remains, within the errors, quite
constant during both bursts and locked to the persistent value (Figure
\ref{fig:burst_ev}, bottom panel). \changes{A detailed spectral and
  temporal analysis of all these type-I bursts was recently reported
  by \citet{Altamirano_10}, although with different techniques.}

\section{Discussion}
Among the AMPs known, \newigrj shows several peculiarities. The source
shows the highest persistent first harmonic fractional amplitude with
a peak value (background corrected) \changes{at the beginning of the
  outburst of about 23\% which linearly decrease to 17\% at the end of
  the outburst (see Figure \ref{fig:amplitudes}), while the highest
  ever observed fractional amplitude was observed by
  \citet{Patruno_10} in \xtejb}.

The pulse shape is complex, showing, on integration times of 3 ks, a
second harmonic, a third harmonic and sporadically a fourth harmonic
with fractional amplitudes of 2.5\%, 1.6\% and 1\%, respectively (see
Figure \ref{fig:amplitudes}).  A third harmonic as strong as the
second harmonic could be, following \cite{Poutanen_06}, an indication
that both hot spots are visible with the secondary spot only partially
visible, since for a single spot the third harmonic should be much
smaller than the second harmonic \citep[$a_2/a_3 \gtrsim
5$,][]{Poutanen_06}, suggesting intermediate values for the
inclination angle. In the AMPs, \citet{Hartman_08} reports of a
sporadically detectable third harmonic in \saxj \changes{, while
  \citet{Patruno_10} detected sporadically a third and a fourth
  harmonic in \xtejb.}

\changes{The fact that in \newigrj the third harmonic is visible for
  nearly all the outburst with a total of 105 detections over 216
  folded pulse profiles makes this source peculiar.}

Moreover, while the fractional amplitude of the first harmonic clearly
shows a steady decrease with the flux, such a decrease is less evident
in the second and third harmonics, for which the fractional amplitude
remains more stable when the X-ray flux decreases (see Figure
\ref{fig:amplitudes}).
\begin{figure*}[ht]
  \subfigure[]{\includegraphics[width=9.0cm]{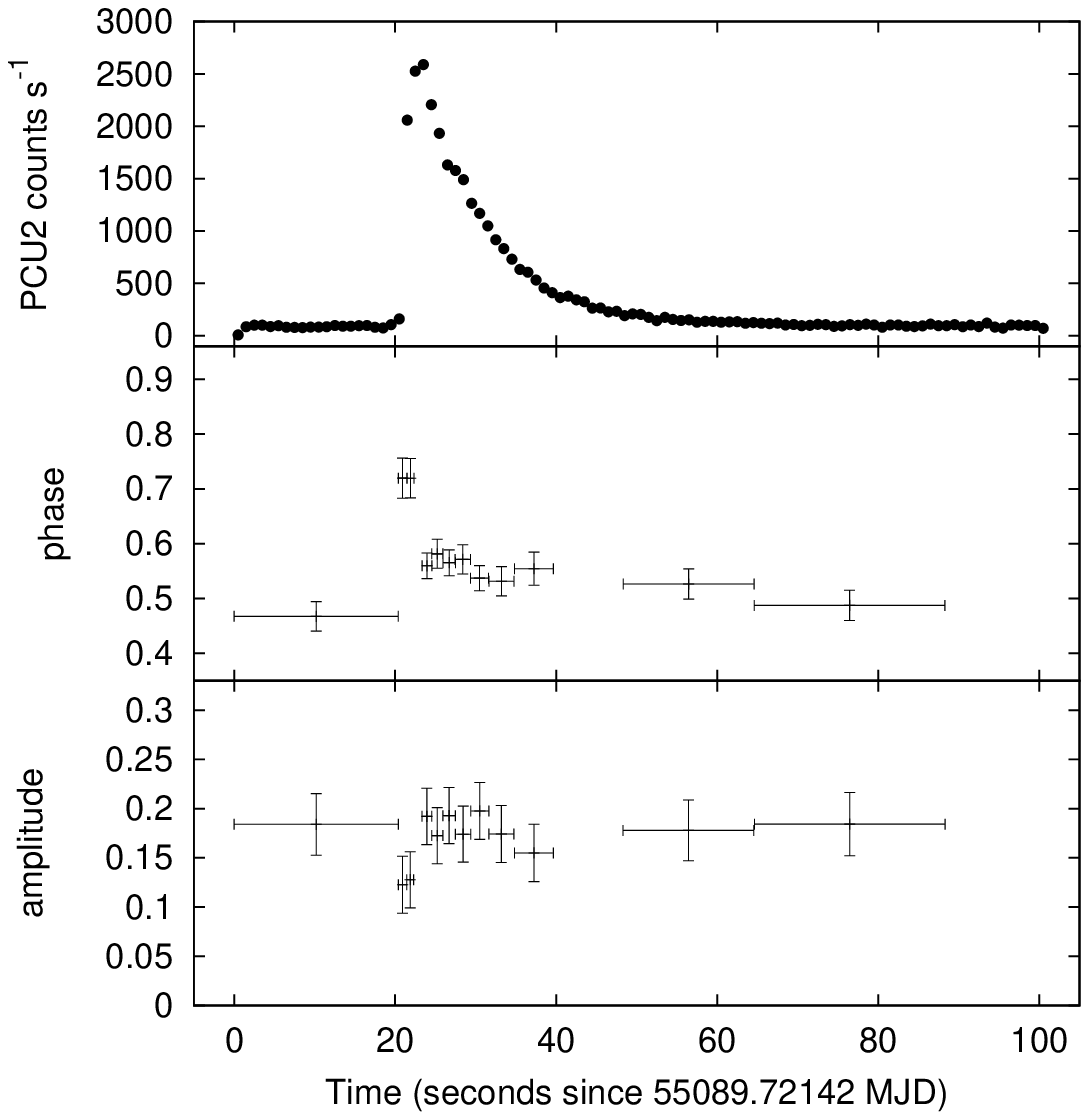}}
  \subfigure[]{\includegraphics[width=9.0cm]{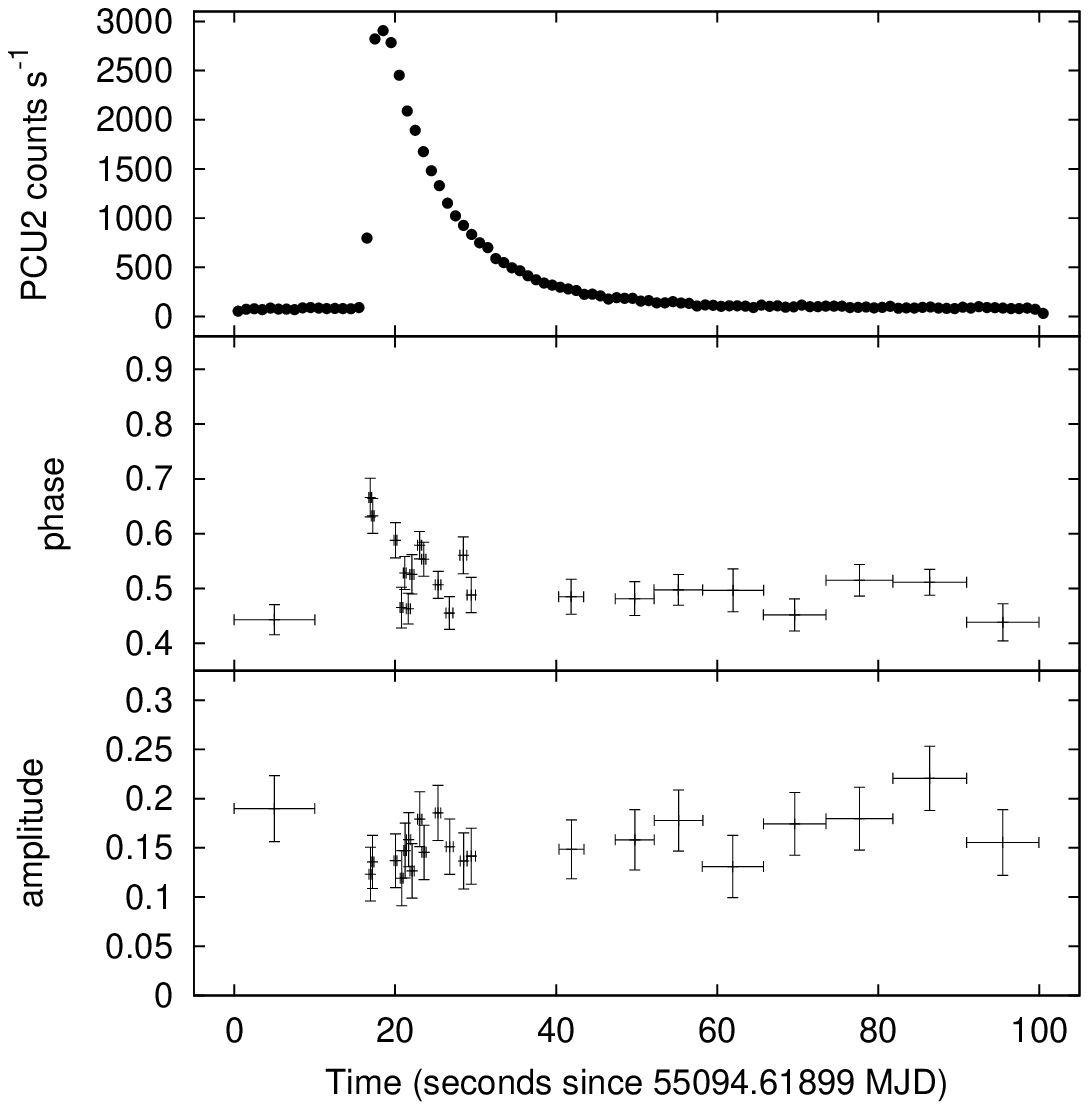}}
  \caption{Two type-I burst timing analysis is reported. For each
    burst in the top panel the PCU2 count rate of the type-I burst
    present in this observation, in the mid panel the pulse phase
    delays and in the bottom panel the fractional amplitude are
    reported.}
  \label{fig:burst_ev}
\end{figure*}

However, the most interesting and puzzling result is the different
behaviour of the phase delays of the four harmonic components. In
particular, similarly to other AMPs, such as \saxj and \xtejb (see
e.g. \citealt{Burderi_07,Hartman_08, Hartman_09,Riggio_08,
  Patruno_10}), the first harmonic shows clear phase fluctuations (see
Figure \ref{fig:first_fit}).
A discussion over the results based on the analysis of the first
harmonic is strongly affected by these phase fluctuations. We can only
do some hypotheses on the nature of such fluctuations. The amplitude
of these fluctuations, $\sim 0.05$ in phase units, could corresponds
to hot spot movements of $\sim 18$ degrees, already seen in numerical
simulations \citep[see][]{Romanova_04, Bachetti_10}, although on
time-scales of fraction of seconds, while the observed time-scales are
of few days.  \changes{We also tried to interpret these phase
  fluctuations with the model suggested by \citet{Patruno_10}. We
  adopted a constant spin frequency model and a linear relation
  between flux and pulse phase residuals. The best correlation
  parameter set gave a $\chi^2_r \simeq 23$, clearly unacceptable.}  A
model to describe and correctly interpret these fluctuations is still
missing.


However, under the working hypothesis that some exchange of angular
momentum between the NS and the accreting matter has to occur during
X-ray outbursts, when the accretion rate is at its maximum, we try to
interpret the behaviour of the phase delays of the second and third
harmonic, which appear to be less affected by phase fluctuations. The
fact that the phase delays derived from the second harmonic appear
more stable than those derived from the first harmonic has been
observed, for instance, during the 2002 outburst from
\saxj. \cite{Burderi_07} showed that, while the first harmonic phase
delays clearly show a phase shift at days 14 from the beginning of the
2002 outburst, a similar phase shift was not present in the phase
delays derived from the second harmonic. A similar behaviour was also
observed by \cite{Riggio_08} for the AMP \xtejb, which went onto
outburst just once in 2003 in the \rxte era.  In both these cases, the
interpretation of the phase delays derived from the second harmonic in
terms of accretion torques provided reasonable spin frequency
derivatives (and inferred mass accretion rates onto the NS),
\changes{although a different interpretation was given for both
  sources by \citet{Hartman_09,Patruno_09}}. Naturally, these results
have to be taken with great caution, since it is not excluded that
phase fluctuations can still affect phase delays derived from the
second (or higher) harmonic.

In the case of \newigrj, again, the second harmonic shows a more
regular behaviour with respect to the first harmonic and suggests a
spin-up of the NS (see Figure \ref{fig:snd_fit}). From the fit with
our simplified torque model we obtain an \Mdot estimate, fixing $\mu =
0$, of \Mdotmax = \np{0.92(10)}{-9} \Msun year$^{-1}$, which would be
compatible with the observed X-ray flux from the source if we put the
source at a distance of $6.3$ kpc. This could indirectly suggest that
the second harmonic component is a better tracer of the spin frequency
evolution, even if a physical model that explains all the
phenomenology observed in all the AMPs class is still missing.
Results obtained on the second harmonic give, as in the case of the
AMPs \saxj \citep{Burderi_06, Hartman_09} and \xtejb
\citep{Riggio_08}, reasonable values for the physical parameters of
the system.

However, if we look at the third harmonic we find that it shows a
spin-up although it is not highly significant
\changes{($\mathbf{\gtrsim 3\sigma}$)} and smaller than the value
inferred from the second harmonic. The lower limit to the accretion
rate \Mdotmax = \np{3.1(9)}{-10} \Msun year$^{-1}$ gives a distance of
3.6(5) kpc, at $\sim 5\sigma$ from the value inferred from the second
harmonic. This value is not compatible with the second harmonic
one. The lack of literature and observations of the third harmonic
behaviour in other AMPs does not allow to make a comparison with other
cases, leaving the question open.

\changes{A fourth harmonic was sporadically detected during the
  outburst. We tried to fit the pulse phase delays with the same
  models adopted for the other harmonics but the result is
  inconclusive. It should be noted that is affected by a phase
  fluctuations ($\sim 0.02$) comparable with the first harmonic's
  fluctuations. It is probable that integrating on long time-scales,
  the large fluctuations make its detection unfeasible.}

We performed a high resolution timing analysis of all the type-I
bursts present in this \rxte observation. The results are very similar
for all the bursts, and we show the burst 2 and 5 (see Figure
\ref{fig:burst_ev}) since these are the best sampled in the \rxte
observation. The first harmonic phase delays appear to rise in
correspondence of the fast rising phase of the burst,
\changes{suggesting a frequency drift during the first few seconds
  since the burst onset \citep{Altamirano_10},} and to return to the
phase value during the persistent emission during the burst decay,
giving some evidence that the burst probably starts not far from the
hot spot in the polar cap. The fractional amplitude in each burst
remains instead locked \changes{(within the errors)} to the persistent
emission value during all the burst, suggesting the quite surprising
fact that the temperature gradient does not vary during the burst.  An
interpretation of a such behaviour is beyond the scope of this work.

From the orbital ephemeris reported in Table \ref{table1} the pulsar
mass function is f$_X$ = \np{1.070854(21)}{-3} \Msun.  From this value
of mass function we can derive a minimum mass for the companion star
of 0.14 \Msun, considering an inclination angle of 90$^o$ and a NS
mass of 1.4 \Msun. With such a minimum mass, the companion star of
\newigrj is one of the more massive companion stars among the AMPs,
together with \xtejb and \saxjb \citep{Altamirano_08}.  Using the
relation $R_{RL_2} = \np{1.2}{10} m^{1/3}_{2, 0.1}
P_{2h}^{2/3}\;\textrm{cm}$ \citep[][]{Paczynski_71}, where $m_{2,
  0.1}$ is the companion mass in 0.1\Msun units and $P_{2h}$ is the
orbital period in two hours units, we obtain for the companion's Roche
lobe radius a value of 0.248 R$_\odot$. Such value is larger than what
is expected for a low mass main sequence star \citep[see][5 Gyr
track]{Chabrier_00}, for which the corresponding radius is about 0.15
R$_\odot$. It can be shown that the contact condition between Roche
lobe and companion star pose a firm lower limit to the inclination of
the system of $\sim 20$ degrees, corresponding to a companion mass of
$\sim 0.45$\Msun.  For smaller inclination angles the companion star
would overfill its Roche lobe. This obviously excludes that the
companion star could be a white dwarf or an helium-core star, while
strongly suggests that the companion star is a main sequence star,
possibly bloated as a consequence of its evolutionary history
\citep{Podsiadlowski_02}. \changes{The nature of the companion star is
  thoroughly discussed in \citet{Papitto_10}}.

\begin{acknowledgements} 
  We thank the anonymous referee for having help us to greatly improve
  the paper.\\
  We also thank Sergey B. Popov for several fruitful discussions.\\
  This work is supported by the Italian Space Agency, ASI-INAF
  I/088/06/0 contract for High Energy Astrophysics.
\end{acknowledgements}

\bibliography{ms}

\begin{thebibliography}{40}
\expandafter\ifx\csname natexlab\endcsname\relax\def\natexlab#1{#1}\fi

\bibitem[{{Altamirano} {et~al.}(2008){Altamirano}, {Casella}, {Patruno},
  {Wijnands}, \& {van der Klis}}]{Altamirano_08}
{Altamirano}, D., {Casella}, P., {Patruno}, A., {Wijnands}, R., \& {van der
  Klis}, M. 2008, \apjl, 674, L45

\bibitem[{{Altamirano} {et~al.}(2010){Altamirano}, {Watts}, {Linares},
  {Markwardt}, {Strohmayer}, \& {Patruno}}]{Altamirano_10}
{Altamirano}, D., {Watts}, A., {Linares}, M., {et~al.} 2010, ArXiv e-prints

\bibitem[{{Bachetti} {et~al.}(2010){Bachetti}, {Romanova}, {Kulkarni},
  {Burderi}, \& {di Salvo}}]{Bachetti_10}
{Bachetti}, M., {Romanova}, M.~M., {Kulkarni}, A., {Burderi}, L., \& {di
  Salvo}, T. 2010, \mnras, 146

\bibitem[{{Baldovin}(2009)}]{Baldovin_ATEL_09}
{Baldovin}, C. 2009, The Astronomer's Telegram, 2196, 1

\bibitem[{{Bevington} \& {Robinson}(2003)}]{Bevington_03}
{Bevington}, P.~R. \& {Robinson}, D.~K. 2003, {Data reduction and error
  analysis for the physical sciences}, 3rd edn. (McGraw-Hill)

\bibitem[{{Bozzo} {et~al.}(2009){Bozzo}, {Ferrigno}, {Falanga}, {Campana},
  {Kennea}, \& {Papitto}}]{Bozzo_09}
{Bozzo}, E., {Ferrigno}, C., {Falanga}, M., {et~al.} 2009, ArXiv e-prints

\bibitem[{{Burderi} {et~al.}(2007){Burderi}, {Di Salvo}, {Lavagetto}, {Menna},
  {Papitto}, {Riggio}, {Iaria}, {D'Antona}, {Robba}, \& {Stella}}]{Burderi_07}
{Burderi}, L., {Di Salvo}, T., {Lavagetto}, G., {et~al.} 2007, \apj, 657, 961

\bibitem[{{Burderi} {et~al.}(2006){Burderi}, {Di Salvo}, {Menna}, {Riggio}, \&
  {Papitto}}]{Burderi_06}
{Burderi}, L., {Di Salvo}, T., {Menna}, M.~T., {Riggio}, A., \& {Papitto}, A.
  2006, \apjl, 653, L133

\bibitem[{{Chabrier} \& {Baraffe}(2000)}]{Chabrier_00}
{Chabrier}, G. \& {Baraffe}, I. 2000, \araa, 38, 337

\bibitem[{{Chenevez} {et~al.}(2009){Chenevez}, {Kuulkers}, {Beckmann}, {Bird},
  {Brandt}, {Domingo}, {Ebisawa}, {Jonker}, {Kretschmar}, {Markwardt},
  {Oosterbroek}, {Paizis}, {Risquez}, {Sanchez-Fernandez}, {Shaw}, \&
  {Wijnands}}]{Chevenez_ATEL_09}
{Chenevez}, J., {Kuulkers}, E., {Beckmann}, V., {et~al.} 2009, The Astronomer's
  Telegram, 2235, 1

\bibitem[{{Deeter} {et~al.}(1981){Deeter}, {Pravdo}, \& {Boynton}}]{Deeter_81}
{Deeter}, J.~E., {Pravdo}, S.~H., \& {Boynton}, P.~E. 1981, \apj, 247, 1003

\bibitem[{{Friedman} \& {Pandharipande}(1981)}]{Friedman_81}
{Friedman}, B. \& {Pandharipande}, V.~R. 1981, Nuclear Physics A, 361, 502

\bibitem[{{Galloway} {et~al.}(2002){Galloway}, {Chakrabarty}, {Morgan}, \&
  {Remillard}}]{Galloway_02}
{Galloway}, D.~K., {Chakrabarty}, D., {Morgan}, E.~H., \& {Remillard}, R.~A.
  2002, \apjl, 576, L137

\bibitem[{{Ghosh} {et~al.}(1977){Ghosh}, {Pethick}, \& {Lamb}}]{Ghosh_77}
{Ghosh}, P., {Pethick}, C.~J., \& {Lamb}, F.~K. 1977, \apj, 217, 578

\bibitem[{{Hartman} {et~al.}(2008){Hartman}, {Patruno}, {Chakrabarty},
  {Kaplan}, {Markwardt}, {Morgan}, {Ray}, {van der Klis}, \&
  {Wijnands}}]{Hartman_08}
{Hartman}, J.~M., {Patruno}, A., {Chakrabarty}, D., {et~al.} 2008, \apj, 675,
  1468

\bibitem[{{Hartman} {et~al.}(2009){Hartman}, {Patruno}, {Chakrabarty},
  {Markwardt}, {Morgan}, {van der Klis}, \& {Wijnands}}]{Hartman_09}
{Hartman}, J.~M., {Patruno}, A., {Chakrabarty}, D., {et~al.} 2009, \apj, 702,
  1673

\bibitem[{{Jahoda} {et~al.}(2006){Jahoda}, {Markwardt}, {Radeva}, {Rots},
  {Stark}, {Swank}, {Strohmayer}, \& {Zhang}}]{Jahoda_06}
{Jahoda}, K., {Markwardt}, C.~B., {Radeva}, Y., {et~al.} 2006, \apjs, 163, 401

\bibitem[{{Leahy} {et~al.}(1983){Leahy}, {Elsner}, \& {Weisskopf}}]{Leahy_83}
{Leahy}, D.~A., {Elsner}, R.~F., \& {Weisskopf}, M.~C. 1983, \apj, 272, 256

\bibitem[{{Markwardt} {et~al.}(2009{\natexlab{a}}){Markwardt}, {Altamirano},
  {Strohmayer}, \& {Swank}}]{Markwardt_ATEL_09b}
{Markwardt}, C.~B., {Altamirano}, D., {Strohmayer}, T.~E., \& {Swank}, J.~H.
  2009{\natexlab{a}}, The Astronomer's Telegram, 2237, 1

\bibitem[{{Markwardt} {et~al.}(2009{\natexlab{b}}){Markwardt}, {Altamirano},
  {Swank}, {Strohmayer}, {Linares}, \& {Pereira}}]{Markwardt_ATEL_09}
{Markwardt}, C.~B., {Altamirano}, D., {Swank}, J.~H., {et~al.}
  2009{\natexlab{b}}, The Astronomer's Telegram, 2197, 1

\bibitem[{{Miller-Jones} {et~al.}(2009){Miller-Jones}, {Russell}, \&
  {Migliari}}]{Miller-Jones_ATEL_09}
{Miller-Jones}, J. C.~A., {Russell}, D.~M., \& {Migliari}, S. 2009, The
  Astronomer's Telegram, 2232, 1

\bibitem[{{Nowak} {et~al.}(2009){Nowak}, {Paizis}, {Wilms}, {Rodriguez},
  {Chaty}, {Ebisawa}, {Del Santo}, {Farinelli}, {Ubertini}, \&
  {Courvoisier}}]{Nowak_ATEL_09}
{Nowak}, M.~A., {Paizis}, A., {Wilms}, J., {et~al.} 2009, The Astronomer's
  Telegram, 2215, 1

\bibitem[{{Paczy{\'n}ski}(1971)}]{Paczynski_71}
{Paczy{\'n}ski}, B. 1971, \araa, 9, 183

\bibitem[{{Pandharipande} \& {Ravenhall}(1989)}]{Pandharipande_89}
{Pandharipande}, V.~R. \& {Ravenhall}, D.~G. 1989, in NATO ASIB Proc. 205:
  Nuclear Matter and Heavy Ion Collisions, ed. {M.~Soyeur, H.~Flocard,
  B.~Tamain, \& M.~Porneuf}, 103--+

\bibitem[{{Papitto} {et~al.}(2008){Papitto}, {Menna}, {Burderi}, {di Salvo}, \&
  {Riggio}}]{Papitto_08}
{Papitto}, A., {Menna}, M.~T., {Burderi}, L., {di Salvo}, T., \& {Riggio}, A.
  2008, \mnras, 383, 411

\bibitem[{{Papitto} {et~al.}(2009){Papitto}, {Riggio}, {Burderi}, {di Salvo},
  {D'A{\`i}}, {Iaria}, \& {Menna}}]{Papitto_ATEL_09}
{Papitto}, A., {Riggio}, A., {Burderi}, L., {et~al.} 2009, The Astronomer's
  Telegram, 2220, 1

\bibitem[{{Papitto} {et~al.}(2010){Papitto}, {Riggio}, {Di Salvo}, {Burderi},
  {D'A{\`i}}, {Iaria}, {Bozzo}, \& {Menna}}]{Papitto_10}
{Papitto}, A., {Riggio}, A., {Di Salvo}, T., {et~al.} 2010, ArXiv e-prints

\bibitem[{{Patruno} {et~al.}(2010){Patruno}, {Hartman}, {Wijnands},
  {Chakrabarty}, \& {van der Klis}}]{Patruno_10}
{Patruno}, A., {Hartman}, J.~M., {Wijnands}, R., {Chakrabarty}, D., \& {van der
  Klis}, M. 2010, \apj, 717, 1253

\bibitem[{{Patruno} {et~al.}(2009){Patruno}, {Wijnands}, \& {van der
  Klis}}]{Patruno_09}
{Patruno}, A., {Wijnands}, R., \& {van der Klis}, M. 2009, \apjl, 698, L60

\bibitem[{{Podsiadlowski} {et~al.}(2002){Podsiadlowski}, {Rappaport}, \&
  {Pfahl}}]{Podsiadlowski_02}
{Podsiadlowski}, P., {Rappaport}, S., \& {Pfahl}, E.~D. 2002, \apj, 565, 1107

\bibitem[{{Poutanen} \& {Beloborodov}(2006)}]{Poutanen_06}
{Poutanen}, J. \& {Beloborodov}, A.~M. 2006, \mnras, 373, 836

\bibitem[{{Rappaport} {et~al.}(2004){Rappaport}, {Fregeau}, \&
  {Spruit}}]{Rappaport_04}
{Rappaport}, S.~A., {Fregeau}, J.~M., \& {Spruit}, H. 2004, \apj, 606, 436

\bibitem[{{Riggio} {et~al.}(2007){Riggio}, {di Salvo}, {Burderi}, {Iaria},
  {Papitto}, {Menna}, \& {Lavagetto}}]{Riggio_07}
{Riggio}, A., {di Salvo}, T., {Burderi}, L., {et~al.} 2007, \mnras, 382, 1751

\bibitem[{{Riggio} {et~al.}(2008){Riggio}, {Di Salvo}, {Burderi}, {Menna},
  {Papitto}, {Iaria}, \& {Lavagetto}}]{Riggio_08}
{Riggio}, A., {Di Salvo}, T., {Burderi}, L., {et~al.} 2008, \apj, 678, 1273

\bibitem[{{Riggio} {et~al.}(2009){Riggio}, {Papitto}, {Burderi}, {di Salvo},
  {D'Ai}, {Iaria}, \& {Menna}}]{Riggio_ATEL_09}
{Riggio}, A., {Papitto}, A., {Burderi}, L., {et~al.} 2009, The Astronomer's
  Telegram, 2221, 1

\bibitem[{{Romanova} {et~al.}(2004){Romanova}, {Ustyugova}, {Koldoba}, \&
  {Lovelace}}]{Romanova_04}
{Romanova}, M.~M., {Ustyugova}, G.~V., {Koldoba}, A.~V., \& {Lovelace},
  R.~V.~E. 2004, \apj, 610, 920

\bibitem[{{Torres} {et~al.}(2009){Torres}, {Jonker}, {Steeghs}, {Simon}, \&
  {Gutowski}}]{Torres_ATEL_09}
{Torres}, M.~A.~P., {Jonker}, P.~G., {Steeghs}, D., {Simon}, J.~D., \&
  {Gutowski}, G. 2009, The Astronomer's Telegram, 2216, 1

\bibitem[{{van den Heuvel}(1984)}]{VanDenHeuvel_84}
{van den Heuvel}, E.~P.~J. 1984, Journal of Astrophysics and Astronomy, 5, 209

\bibitem[{{Zdziarski} {et~al.}(1996){Zdziarski}, {Johnson}, \&
  {Magdziarz}}]{Zdziarski_96}
{Zdziarski}, A.~A., {Johnson}, W.~N., \& {Magdziarz}, P. 1996, \mnras, 283, 193

\bibitem[{{{\.Z}ycki} {et~al.}(1999){{\.Z}ycki}, {Done}, \& {Smith}}]{Zycki_99}
{{\.Z}ycki}, P.~T., {Done}, C., \& {Smith}, D.~A. 1999, \mnras, 309, 561

\end{thebibliography}

\end{document}